\documentclass[11pt]{amsart}
\pdfoutput=1

\usepackage[utf8]{inputenc}

\usepackage[margin=3.25cm]{geometry}

\usepackage{amssymb, amsmath, amsthm, amsfonts}
\usepackage{graphicx}
\usepackage{color}
\usepackage{mathrsfs}
\usepackage{physics}
\usepackage{mathtools}
\usepackage{tikz}
\usepackage{faktor}
\usepackage{tikz-cd}
\usepackage{bm}
\usepackage[hidelinks]{hyperref}
\usepackage{csquotes}

\DeclareRobustCommand{\SkipTocEntry}[5]{}

\newtheorem{theorem}{Theorem}
\newtheorem{lemma}{Lemma}
\newtheorem{corollary}{Corollary}

\newtheorem{prop}{Proposition}

\theoremstyle{remark}
\newtheorem{remark}{Remark}
\theoremstyle{remark}

\usepackage[UKenglish]{isodate}
\usepackage[UKenglish]{babel}

\newcommand{\R}{\mathbb{R}}

\DeclareMathOperator{\E}{\mathbb{E}}

\renewcommand{\grad}{\nabla}

\newcommand{\Hom}{\mathrm{Hom}}

\newcommand{\T}{T^0_{\text{inf}}}
\newcommand{\A}{\mathcal{A}}
\newcommand{\Cob}{\mathrm{Cob}}
\newcommand{\pss}{p_{\mathrm{ss}}}

\newcommand{\Obj}{\mathrm{Obj}}
\newcommand{\Mor}{\mathrm{Mor}}
\newcommand{\Diff}{\mathrm{Diff}}
\newcommand{\D}{\mathscr{D}}

\title[Functorial statistical physics]{Functorial statistical physics: Feynman--Kac formulae and information geometries}

\author{Dalton A R Sakthivadivel}
\address{\parbox{\linewidth-12pt}{Department of Mathematics, Department of Physics and Astronomy, Stony Brook University, Stony Brook, NY, 11794-3651, USA}}
\address{VERSES Research Lab, Los Angeles, CA, 90016, USA}
\email{dalton.sakthivadivel@stonybrook.edu}
\urladdr{https://darsakthi.github.io}

\date{\today}
\subjclass[2020]{Primary 53B12, 82C05; Secondary 46M20, 46T12}

\let\oldtocsection=\tocsection

\let\oldtocsubsection=\tocsubsection

\renewcommand{\tocsection}[2]{\hspace{0em}\oldtocsection{#1}{#2}}
\renewcommand{\tocsubsection}[2]{\hspace{19.5pt}\oldtocsubsection{#1}{#2}}

\begin{document}

\maketitle

\begin{abstract}

The main results of this paper comprise proofs of the following two related facts: (i) the Feynman--Kac formula is a functor $F_*$, namely, between a stochastic differential equation and a dynamical system on a statistical manifold, and (ii) a statistical manifold is a sheaf generated by this functor with a canonical gluing condition. Using a particular locality property for $F_*$, recognised from functorial quantum field theory as a `sewing law,' we then extend our results to the Chapman--Kolmogorov equation {\it via} a time-dependent generalisation of the principle of maximum entropy. This yields a partial formalisation of a variational principle which takes us beyond Feynman--Kac measures driven by Wiener laws. Our construction offers a robust glimpse at a deeper theory which we argue re-imagines time-dependent statistical physics and information geometry alike. 

\end{abstract}

%%%%%%%%%%%%%%%%%%%%%%%%%%%%%%%%%%%%%%%%%%%%%%%%%%%%%%%%%%%%%%%%%%%%%

\section{Preliminaries}

We will begin by taking a finite-dimensional manifold $X$ globally equivalent to $\R^d$ or some submanifold thereof, and its diffeomorphism group $\Diff(X)$. A smooth dynamical system on $X$ is a pair $(\varphi_0, v)$, where $v$ is a map from $X \to TX$ ({\it i.e.}, a vector field) and $\varphi_0$ is an initial point $x_0$. Under well-known regularity conditions on $(\varphi_0, v)$ this gives rise to a unique orbit, or a one-parameter family of local diffeomorphisms $\varphi : X \cross I \to X$ . The map $\varphi_t$ is thus a path in $\Diff(X)$, and since this is a one-parameter group we have the equations $\varphi_s \circ \varphi_t = \varphi_{t+s}$ and so forth. Hence, a dynamical system regarded this way is a category generated by the set of endomorphisms of $X$. In this paper, we are interested in a certain functorial relationship between dynamical systems in different spaces; in particular, this paper is motivated by the fact that such a functor represents a map between complicated partial differential equations for diffusion and the simpler time evolution of the parameters of their stochastic representations. 

For our purposes, a probability density is a smooth function on a continuous support $M$ which can be identified as the Radon--Nikodym derivative of some probability measure $\dd{P(q)} = p(q) \dd{q}$ on a manifold $M$. It is possible to define a {\it statistical manifold on $M$}, $S_M$, as a finite-dimensional (unless otherwise specified) manifold of probability densities parameterised by some unique vector in $\R^n$ \cite{lauritzen1987statistical, cena}. It is further possible to place a particular Riemannian structure on such a space, called an information geometry \cite{amari}. In particular, we will consider exponential probability densities, making points in $\mathrm{Obj}(S_M)$ probability densities (up to a normalisation constant)
\[
p(q) = \exp{-c_\lambda J(q)},
\]
where $J(q)$ is a scalar function $J : M \to \R$, $q \mapsto h$, and $\lambda$ is the aforementioned $n$-dimensional vector. More generally, we will consider sums of these quantities, 
\[
\exp{-\sum_i^n c_{\lambda, i} J_i(q)},
\]
and will denote this as the scalar product of vectors $\lambda \cdot J$. Trajectories in $S_M$ are one-parameter groups 
\[
p(q, t) = \exp{-c_\lambda(t) \cdot J(q, t)}
\]
with $J : (q, t) \mapsto h(t)$ which assigns a time-dependent scalar quantity to every state $q$, such as (for example) $L_2$ distance away from a non-stationary mean.

A good example describes the Wiener process
\begin{equation}\label{wiener-proc}
\dd{Y_t} = -b\dd{t} + \sqrt{2\lambda}\dd{W_t}.
\end{equation}
Let $\pi$ be the Wiener measure describing sample paths of the process $W_t$. By It\=o's lemma, the moments of the probability distribution $p(q, t)$ are 
\[
\E_\pi[Y_t] = -\int_0^t b\dd{s} = -bt, \qquad \E_\pi\big[Y_t^2 - \E[Y_t]^2\big] = \int_0^t \lambda \dd{s} = \lambda t.
\] 
and the non-stationary solution to the Fokker-Planck equation for \eqref{wiener-proc},
\[
\pdv{t} p(q, t) = -b\pdv{q} p(q, t) + \lambda \pdv[2]{}{q} p(q, t),
\]
with initial condition $p(q, t) = \delta(q - q_0)$, is
\[
\frac{1}{2\sqrt{\pi \lambda t}}\exp{-\frac{(q -q_0 - bt)^2}{4\lambda t}}
\]
Here, it is evident that $\lambda$ scales noise amplitude\textemdash and thus variance\textemdash in a way that is formalised by the notion of a `speed parameter' in large deviations theory. Moreover, it is clear that the two degrees of freedom shaping $p(q, \tau)$ at a fixed time $\tau$ are $b\tau$ and $\lambda \tau$, which are the mean and the variance at $\tau$, respectively.

Let us also consider the case of an Ornstein--Uhlenbeck process
\[
\dd{Y_t} = -b(Y_t-k)\dd{t} + \sqrt{2\lambda}\dd{W_t}.
\]
Differentiating under the integral, It\=o's lemma yields a set of unidirectionally coupled ordinary differential equations for moments, 
\begin{gather*}
\dot m(t) = \dv{t}\E_\pi[Y_t] = -b\E[Y_t] + k\\
\dot s(t) = \dv{t} \E_\pi\big[Y_t^2 - m(t)^2\big] = -2b\E_\pi\big[Y_t^2\big] + 2\lambda - \partial_t (m(t)^2),
\end{gather*}
such that
\[
m(t) = e^{-bt}m_0 + k\left(1 - e^{-bt}\right),
\]
and after a change in variables $s_t \mapsto s_t + m_t^2$, 
\[
s(t) = s_0e^{-2bt} + \frac{\lambda}{b}\left( 1 - e^{-2bt} \right) - m(t)^2.
\]
Given initial conditions $m_0$ and $s_0$, these form a dynamical system $(m_0 \cross s_0, b \cross \sigma^2)$ with trajectory $\varphi(t) = (m(t), s(t))$ valued in $\R^2$. In the limit $t \to \infty$ we obtain the stationary mean $k$ (indeed, we describe the Ornstein--Uhlenbeck process as mean-reverting). The variance also stabilises in the infinite time limit, meaning that the equilibrium solution to the Ornstein--Uhlenbeck process is parameterised by two values: the constants for the mean and the variance.\footnote{Here one can derive that $\lambda = \sigma^2/2$, such that $1/4\lambda = 1/2\sigma^2$.} This yields 
\[
p(q, t \mid q_0, t_0) \propto \exp{-\frac{(q - m(t))^2}{2s(t) - s_0e^{-2bt}}}
\]
with equilibrium
\[
p(q) \propto \exp{-\frac{b(q-k)^2}{2\lambda}}.
\]
Then, at steady state, the two parameters are the mean (zero) and the variance ($b/\lambda$). Note how changing whether $b$ is a mean or a variance completely changes the SDE. Note also that the equilibrium solution maximises entropy under the constraining potential function $\grad J(q) = b(q-k)$, indeed implying $\E[Y_t] = k$ at steady state.

More generally, suppose $\dim(M) = d$. Take a non-stationary stochastic process on $M$ driven by an $\ell$-dimensional Wiener process, {\it i.e.},
\begin{equation}\label{sde-eq}
\dd{Y^i_t} = b_i(Y_t, t)\dd{t} + \sigma_{ik}(Y_t, t) \dd{W^k_t}
\end{equation}
in the It\=o picture, $i \in \{1, \ldots, d\}$ and $k\in \{1, \ldots, \ell\}$. Given the boundary condition of a final state $p(q, T) = \tilde p(q)$, the stochastic differential equation has a probability measure described for all $t \leq T$ by the nonlinear Kolmogorov backwards equation
\[
\pdv{t} p(q, t) = -\sum^d_i b_i(q, t)\pdv{q^i}p(q, t) - \frac{1}{2}\sum_{i}^d\sum_{j}^d\left(\sum_k^\ell \sigma_{ik}(q,t)\sigma_{jk}(q,t) \pdv{}{q^i}{q^j} p(q, t)\right).
\]
It is well-known that this equation is solved by the Feynman--Kac formula
\[
p(q, t) = \E_{\pi}\left[\, p(Y_T, T) \; \middle| \; Y_t = q_t \right]
\]
where $\pi$ is the law for $\dd{W_t}$ ({\it i.e.}, the Wiener measure on sample paths $W_t$), $t$ is in $[0, T]$, $Y_t = q_t$ is an initial condition for \eqref{sde-eq}, and $p(Y_T, T)$ is a terminal state for the diffusion (for instance, a known or desired steady state measure).\footnote{Note that we could apply the same analysis to the Kolmogorov forwards equation, also called the Fokker--Planck equation.}

From this reasoning it is clear that there is some sort of, perhaps rather involved, map between a set of ODEs parameterising an SDE and a PDE describing that SDE\textemdash a map which is made possible by the common language of the moments or {\it sufficient statistics} of an SDE. Moreover, if these parameters are time-dependent functions, that ODE induces a vector field on a manifold of parameterised probability distributions, mapping $(m_0 \cross s_0, \dot m \cross \dot s)$ to $(p_0, \partial_t p)$. In particular, fix $J(q)$. It follows that a pair $(m, \lambda)$ uniquely represents a $p(q)$; likewise, a change in $c_\lambda$ ought to uniquely determine a change in $p(q)$. Hence a function 
\[
f : (m_1, \lambda_1) \to (m_2, \lambda_2)
\]
induces a map 
\[
g : \exp{-c_{\lambda,1} J(q)} \to \exp{-c_{\lambda,2} J(q)}
\]
in a way that also has some geometric content (in the sense that it ought to be a path on $S_M$ as well). 

The aim of this paper is two-fold. To begin, we wish to formalise the intuition that statistical manifolds are generated by a functor relating a change in parameters to a change in points on the manifold. The fact that operations on the statistical manifold push forward along that functor justifies the usual appeals to the parameter space one encounters in this literature. To achieve this we propose a new characterisation of the jet groupoid.

Though of independent interest, this can be seen as a technical result supporting the transfer of path-dependent methods in stochastic analysis to the study of time-dependent statistical physics. In areas of time-dependent statistical mechanics (like the study of systems far from equilibrium) we are often interested in dynamical systems whose sampling statistics\textemdash that is, the parameters of their probability densities\textemdash change. We wish to apply the viewpoint introduced above to see how productive it might be in this area. Using the construction here we are able\textemdash in particular its functorial properties\textemdash we are able to recover Jaynes' principle of maximum calibre, a path-integral-based variational principle generalising the entropy function to time-dependent processes. 

Altogether, this construction follows along the intuitive idea that the `material' physics of a system can be lifted to the statistical properties of that system, and that calculations and quantities in one space reflect those in the other. It also suggests new ways of achieving this lift in such a manner as to complement existing approaches to statistical physics. Hence, it illuminates new and existing techniques with which to handle random dynamical systems in non-equilibrium statistical physics. 

\section{A functorial perspective on the Feynman--Kac formula}

We consider smooth curves defined locally on $X$ as trajectories of the dynamical system mentioned above. Since any such $\varphi_t$ is a family of $C^\infty$-local diffeomorphism mapping $X$ to $X$, it can be taken in the Lie group which deloops $\Diff_\ell^\infty(X)$ to the category $\D$ where $\Obj(\D) = X$, taken as a one object set, and $\Mor(\D) = \Hom(X, X)$ with appropriate conditions on plots. From there, the order-zero jet groupoid $\Gamma_0(X)$, whose morphisms are local diffeomorphisms between points $x_1$ to $x_2$ in $X$, can be formed. 

\begin{prop}\label{path-groupoid-prop}
    If points in $X$ parameterise points in $S_M$, then there exists an isomorphism of groupoids taking paths in $X$ to paths in $S_M$ in such a way that an orbit of probabilities is induced on $S_M$.
\end{prop}
\begin{proof}
    Suppose every $x$ maps uniquely to a $p(q)$ via the assignment $x \mapsto F(x)$, 
    \[
    F(x) = \exp{-x\cdot J(q)}, \quad x \in \R^n.
    \]
    It is easy to verify that this is a continuous bijection on $S_M$ for fixed $J(q)$. This pushes forward to a homomorphism of jet groupoids by  
    \[
    F_* : \Gamma_0(X) \to \Gamma_0(S_M),
    \]
    where $F_* = F\circ \varphi$ for any local diffeomorphism $\varphi : x_1 \to x_2$. This map is moreover an isomorphism by the uniqueness of parameters\textemdash which, specifically, provides invertibility of $F_*$.
\end{proof}

Note in particular that a homomorphism of groupoids is a functor. Since elements of $\Gamma_0(S_M)$ are diffeomorphisms of $S_M$, they ought to be given as the orbit of some dynamical system when $S_M$ is equipped with its tangent bundle $TS_M$. We now want some way to achieve a differential of $F_*$: a map that lifts tangent vector fields\textemdash infinitesimal paths in $\Diff(X)$\textemdash from the Lie algebroid of $\Gamma_0(X)$ to the Lie algebroid of $\Gamma_0(S_M)$. This is possible via the definition of the Atiyah Lie algebroid of $\Gamma_0(-)$, which lifts morphisms in $\Gamma_0(-)$ to isomorphisms of fibres of infinitesimal paths over a local neighbourhood, thereby inducing such a vector field. The differential $\dd{f} : T_e\Diff_\ell^\infty(X) \to T_e\Diff_\ell^\infty(S_M)$ categorifies to 
\[
\dd{F_*} : T^0_{\text{inf}}X \to T^0_{\text{inf}}S_M
\]
where $T^0_{\text{inf}}$ constructs the Atiyah Lie algebroid for $\Gamma_0$.

\begin{prop}\label{differential-prop}
    Suppose motion on $S_M$ is given by the Kolmogorov backwards equation. The differential of $F_*$ can be written as an assignment of tangent vectors in $X$ to the action of infinitesimal generators on $S_M$. 
\end{prop}
\begin{proof}
    Recall that a representation of an algebra $\A$ on a space $N$ is a map $\rho : \A \to \Hom(N, N)$ assigning elements $A \in \A$ to endomorphisms of $N$. Under the representation $\rho A = -A\cdot-$ such that $\partial_t p = \rho A$, we have
    \[
    \begin{tikzcd}
    &\A \arrow[d, "\rho"] \\
    \T X \arrow[r, swap, "\dd{F_*}"] \arrow[ur, dotted, "R"] & \T S_M,
    \end{tikzcd}
    \]
    given by $\dd{F_*} = \rho \circ R: \big(
    \partial_t m(q, t), 
    \partial_t s(q,t)\big) \mapsto A \mapsto -Ap$ with $A = m(q,t) \partial_q + s(q, t) \partial_{qq}$. Hence each Hom set in the image of $\dd{F_*}$ can be assigned to an operator under the representation of $\A$ by left multiplication of $A \in \A$ with points $Fx_i$. From this we obtain 
    \begin{equation}\label{diff-functor}
    \dd{F_*} : \T X \to \rho \A.
    \end{equation}
\end{proof}

\begin{remark}
    Note that $A$ is indeed precisely a tangent vector to $\Diff$, when double derivatives are introduced.
\end{remark}

\begin{prop}\label{integral-prop}
    Let $\mathcal{P}$ be a group of Markov kernels acting on $S_M$, such that for $p(q, T) = \pss(x)$ and some $P_t \in \mathcal{P}$, any $p(q,t) = P_t\pss(x)$, $t \leq T$. There exists a formal integral for \eqref{diff-functor}, 
    \[
    F_* : \Gamma_0(X) \to \rho\mathcal{P}.
    \]
    Moreover, $F_*$ is a monoidal functor that assigns paths in $\Gamma_0(X)$ to Markov kernels.
\end{prop}
\begin{proof}
    Take Lie integration as a morphism of Hom sets and construct $L : \dd{F_*} \to F_*$.
    By a theorem of Dynkin it is known that $Ap$ is the time derivative of the curve defined by $t \mapsto P_t \pss(x)$. The representation $\rho : \mathcal{P} \to \Hom_{S_M}(Fx_1, Fx_2)$ is now left multiplication by a Markov kernel with infinitesimal generator $A$. By construction, the time evolution of any density $Fx$ is contained in the time evolution of its parameters. In particular, $F(\varphi)$ is a map $F(x_1) \to F(x_2)$ for any $\varphi : x_1 \to x_2$. Concretely, this map behaves as follows: $F( \varphi : x_1 \mapsto x_2)$ is mapped to a change in statistics $p(q; x_1) \mapsto p(q; x_2)$, and the isomorphism takes the propagator $p(q; x_2 \mid x_1)$ to the action of a chain of kernels $P_2 \circ \ldots \circ P_1$ on $p(q; x_1)$. Finally, recall that $F_*$ respects the composition of paths. Hence, $F_*$ is monoidal.
\end{proof}

This yields that, if $A$ is the formal tangent vector of some time-dependent $p(q, t)$ (where the time-dependence is contained in the parameters as a matter of definition) then $F_*$ takes values in the space of actions of Markov kernels $P_t$. The Dynkin formula is crucial to generating a vector field $(p_0, \partial_t p)$ out of $\varphi$.

Interestingly, both $\T X \to \T S_M$ and $\T X \to \Gamma_0(S_M)$ factor through $\Gamma_0(X)$; this corresponds to the intuition that having or solving dynamics on $S_M$ requires knowing the statistics of those dynamics. We will see how this appears in an explicit form for $F_*$.

\begin{theorem}
    The functor $F_*$ is the Feynman--Kac formula.
\end{theorem}
\begin{proof}
    By Dynkin's formula, $P_t \pss = \E^q[\pss(Y_t)]$ satisfies $\partial_t p = Ap$, where $Ap = \dd{F\varphi'}$ by Propositions \ref{differential-prop} and \ref{integral-prop}. This is easily recognised as the Feynman--Kac formula
    \[
    \E_\pi[p(Y_t) \mid Y_0 = q]
    \]
    under the filtration so defined. By It\=o's lemma, this is given in turn by
    \[
    \E_\pi\left[p\left(\E[Y_t] + \sqrt{\mathbb{V}(Y_t)}\right) \mid Y_0 = q\right],
    \]
    making $F_*$ equivalent to 
    \[
    (m_t, s_t) \mapsto \mathrm{FK}(m_t, s_t) = p(q, t).
    \]
\end{proof}

\begin{lemma}
    Let $G$ be a groupoid internal to $\mathsf{DiffSp}$. Oppositisation preserves diffeological structure.
\end{lemma}
\begin{proof}
    Oppositisation can be taken as a base change for the sheaf formed by the diffeological structure on $G$. Since $G^{\mathrm{op}}$ is an isomorphism of the base, the diffeological structure is preserved.
\end{proof}

\begin{theorem}\label{functor-thm}
    If points in $X$ parameterise points in $S_M$, then for $\Gamma_0(X)$ and $\Gamma_0(S_M)$ taken as small categories, the points and paths in a statistical manifold form a $\Gamma_0(S_M)$-valued presheaf on $\Gamma_0(X)$ under a particular pullback. As such, a statistical manifold is generated from a space of parameters $X$ as a sheaf over $X$. Moreover, there exists a cosheaf given by the copresheaf corresponding to the covariant case.
\end{theorem}
\begin{proof}
Take $X$ and $S_M$ as described above. By Proposition \ref{path-groupoid-prop}, there exists a full and faithful functor $F_* : \Gamma_0(X) \to \Gamma_0(S_M)$ induced by the map $F : X \to S_M$ on objects. Taking the precomposition of $F$ with oppositisation of $\Gamma_0(X)^{\mathrm{op}}$, we obtain 
\[
\tilde F_* : \Gamma_0(X)^{\mathrm{op}} \to \Gamma_0(S_M).
\]
To prove $\tilde F_*$ is a sheaf, simply note that $\Gamma_0(-)$ is a Lie groupoid and hence trivially a diffeological space, as is its opposite, and thus that the presheaf of sets underlying $\tilde F_*$ is a sheaf; from this it follows that the required compatibility conditions on local charts are necessarily satisfied by $\tilde F_*$. The corresponding cosheaf for $F_*$ can be produced by postcomposition of $\tilde F_*$ with oppositisation of $\Gamma_0(S_M)$, for which the same argument holds.
\end{proof}

Theorem \ref{functor-thm} licences a rigorous interpretation of a statistical manifold as an object that keeps track of how parameters appear in probability densities $F(x)$, and how a given parameter $x'$ arises from paths from any other parameter $x$, ({\it i.e.}, $F(\varphi)$ keeps track of sub-objects of $x'$). Sheafifying $\tilde F_*$ has the following intuition: this rule glues together around points in such a way that it assembles into a geometric space, wherein $F(x)$ becomes a `generalised point.' This is almost a triviality, since $\tilde F_*$ respects the diffeological structure of image and preimage. The same preservation of gluing applies to the covariant construction of a cosheaf out of $F_*$ as well.

\begin{remark}
We posit this explains the effectiveness of conventional approaches to information geometry: the manipulation of quantities in $S_M$ is functorially related to calculations in the parameter space $X$, and so an operation on parameters pushes forward to $S_M$. Moreover, this functor is such that the basic essence of a space (and specifically a differentiable space) is preserved. This allows statisticians to ignore the actual geometry of a statistical manifold and simply work with the dynamics on $S_M$, as yielded by the functor we describe here; hence our claim that information geometry in practice is, in some suitable sense, really the study of a particular Riemannian functorial field theory (to be formalised in the following results).
\end{remark}

Essentially we have proven that the Feynman--Kac formula is a recipe for relating the time evolution of the statistics of a non-stationary process to the time evolution of the probability density associated to its samples. Whilst not a surprise, what is interesting is that it is a functorial relationship, whose diffeological nature generates a statistical manifold as a sheaf associated to such an abstract relationship.

The following is not proven, as it is an immediate consequence of the invertibility of $F_*$.

\begin{corollary}
Let $F^{-1} = G$. Proposition \ref{path-groupoid-prop} and Theorem \ref{functor-thm} imply the existence of adjoint functors
\[
G_* : \Gamma_0(S_M) \to \Gamma_0(X)
\]
and
\[
\tilde G_* : \Gamma_0(S_M)^{\mathrm{op}} \to \Gamma_0(X),
\]
and thus any quantities computed thereof.
\end{corollary}

\section{Maximisation of time-dependent entropy from gluing}

The Feynman--Kac measure can be regarded as a Riemannian field theory for Markov kernels, in the sense described here. The notion of a functorial quantum field theory\footnote{We recommend Stolz, \emph{Functorial Field Theories and Factorization Algebras}, for an excellent review\textemdash one with a particular eye towards Riemannian field theories.} as we refer to it is suggested in \cite{schreiber2009aqft} (based on work of Atiyah \cite{atiyah1988topological} and Segal \cite{segal1988definition}, as well as Kontsevitch) and refined in key papers like \cite{freed} and \cite{grady}; in particular, it is construed as a functor from $n$-dimensional bordisms (with some geometric structure $\mathfrak{R}$) to the category of vector spaces and isomorphisms between them $\textsf{Vect}_{\mathrm{isos}}$, which pulls back to a functor from the path $k$-groupoid to $\textsf{Vect}_{\mathrm{isos}}$. Here the path $k$-groupoid mirrors the structure of $\Gamma_k(X)$, in the sense that it is a diffeological groupoid whose coherence condition for higher morphisms is simply that any $(i+1)$-form pulled back to a thin $(i+1)$-morphism has rank of at most $i$, $0 < i \leq k$. The case for Riemannian structure has perhaps been primarily studied by \cite{kandel} with noteworthy results also found in \cite{pickrell}, \cite{grenier2019reflection}, and \cite[\textsection 6]{grady}, as well as the work of Costello and Gwilliam.

\begin{lemma}
    There is an equivalence of categories between $\Gamma_0(X)$ and $1\Cob_n$, as well as between $\Gamma_0(S_M)$ and $\mathsf{Vect}_{\mathrm{isos}}$.
\end{lemma}
\begin{proof}
Morphisms in $1\Cob_n$ are diffeomorphisms $Y$ between $n$-dimensional manifolds $X$. Likewise, $S_M$ is a finite-dimensional Hilbert space, and morphisms in $\mathsf{Vect}_{\mathrm{isos}}$ are invertible maps between vector spaces.
\end{proof}

\begin{corollary}\label{fqft-corollary}
The functor $F_*$ yields a one-dimensional TQFT after identifying $\Gamma_0(X)$ with $1\Cob_n$ and $\Gamma_0(S_M)$ with $\mathsf{Vect}_{\mathrm{isos}}$. Placing a Riemannian structure on $X$ recovers the fact that the Feynman--Kac formula is a Riemannian functorial field theory.
\end{corollary}

We may now comment on Feynman--Kac measures. Here, $P_t$ plays the role of a transition amplitude within a propagator. Indeed, we can deduce 
\begin{equation}\label{prop-eq}
\bra{x} P_tp \ket{x'} = \E[\pss(q; x(t, t'))]
\end{equation}
where the right-hand side is a propagator from incoming states to outgoing states and the left-hand side is a path integral. From this, according to now well-known procedures, we can derive the sewing law characteristic of functorial QFT as construed by Atiyah--Kontsevitch--Segal and which is centred in \cite{schreiber2009aqft}. 

\begin{remark}
In the time-dependent case, we are not interested in operators yielded by propagators acting on states, we are interested in changes in statistics that give the measure a time-dependence. If the parameter is a position or state vector, this becomes equivalent to a conventional functorial QFT via the analytically-continued Feynman path integral. More generally, if $p$ is a measure of Gibbs-type at each time-point then this is standard thermal quantum field theory via the Boltzmann transition weight (also called the Matsubara formalism).
\end{remark}

The desire to move beyond processes driven by a Wiener law remains. If we take the idea seriously, we should look for something like a path integral or else some variational principle that yields a measure on the space of paths consistent with the gluing axioms we have deduced here; something that covers the more general Chapman--Kolmogorov equation.

\begin{theorem}
    Via the Atiyah--Kontsevitch--Segal gluing axioms, Jaynes' principle of maximum calibre is a functorial Euclidean field theory.
\end{theorem}
\begin{proof}
    By Corollary \ref{fqft-corollary}, we have the gluing law 
    \[
    \bra{x} P_tp \ket{x'} = \E[\pss(q; x(t, t'))]
    \]
    under the Markov kernel $P_t^\pi$. We want a more general 
    \[
    \bra{\varphi_0} P_tp \ket{\varphi_s} = p(q(t); \varphi(s))
    \]
    where $p(q(t); \varphi(t))$ is identified with a space-time-dependent measure $p(q, t)$ and $\varphi(t)$ is a time-dependent parameter. Suppose each $P_t$ is a measure of Gibbs-type with infinitesimal generator given by an arbitrary transition matrix. Then we have 
    \[
    e^{-\lambda_f J(t_f)}\circ \ldots \circ e^{-\lambda_2 J(t_2)} \circ e^{-\lambda_1 J(t_1)}.
    \]
    under the AKS gluing law. Such a measure is known to maximise the calibre functional introduced by Jaynes.
\end{proof}

See \cite{jaynes1985macroscopic, dill} for details. Like AKS did with Feynman's path integral, we postulate that\textemdash if formalised\textemdash the thing that the calibre functional `ought' to be is an algebraic object that glues in the way prescribed above. This raises further questions about the analytic nature of certain time-dependent approaches to statistical physics which are not driven by Wiener processes. We hope to investigate them in the future.

\bibliographystyle{alpha}
\bibliography{main}

\begin{thebibliography}{PGLD13}

\bibitem[Ama16]{amari}
Shun-ichi Amari.
\newblock {\em Information Geometry and its Applications}, volume 194 of {\em
  Applied Mathematical Sciences}.
\newblock Springer, 2016.

\bibitem[Ati88]{atiyah1988topological}
Sir Michael~F Atiyah.
\newblock Topological quantum field theory.
\newblock {\em Publications Math{\'e}matiques de l'IH{\'E}S}, 68:175--186,
  1988.

\bibitem[FH21]{freed}
Daniel~S Freed and Michael~J Hopkins.
\newblock Reflection positivity and invertible topological phases.
\newblock {\em Geometry \& Topology}, 25(3):1165--1330, 2021.

\bibitem[GP21]{grady}
Daniel Grady and Dmitri Pavlov.
\newblock The geometric cobordism hypothesis.
\newblock 2021.
\newblock Preprint arXiv:2111.01095.

\bibitem[Gre19]{grenier2019reflection}
Joseph~W Grenier.
\newblock {\em Reflection Positivity: A Quantum Field Theory Connection}.
\newblock 2019.
\newblock PhD thesis.

\bibitem[Jay85]{jaynes1985macroscopic}
Edwin~T Jaynes.
\newblock Macroscopic prediction.
\newblock In {\em Complex Systems—Operational Approaches in Neurobiology,
  Physics, and Computers}, pages 254--269. Springer, 1985.

\bibitem[Kan16]{kandel}
Santosh Kandel.
\newblock Functorial quantum field theory in the {R}iemannian setting.
\newblock {\em Advances in Theoretical and Mathematical Physics},
  20(6):1443--1471, 2016.

\bibitem[Lau87]{lauritzen1987statistical}
Stefan~L Lauritzen.
\newblock Statistical manifolds.
\newblock In {\em Differential Geometry in Statistical Inference}, volume~10 of
  {\em Institute of Mathematical Statistics Lecture Notes}. 1987.

\bibitem[PGLD13]{dill}
Steve Press\'e, Kingshuk Ghosh, Julian Lee, and Ken~A Dill.
\newblock Principles of maximum entropy and maximum caliber in statistical
  physics.
\newblock {\em Reviews of Modern Physics}, 85(3):1115--1141, Jul 2013.

\bibitem[Pic08]{pickrell}
Douglas~M Pickrell.
\newblock ${P}(\phi)_2$ quantum field theories and {S}egal's axioms.
\newblock {\em Communications in Mathematical Physics}, 280(2):403--425, 2008.

\bibitem[PR99]{cena}
Giovanni Pistone and Maria~Piera Rogantin.
\newblock The exponential statistical manifold: mean parameters, orthogonality
  and space transformations.
\newblock {\em Bernoulli}, 5(4):721--760, 1999.

\bibitem[Sch09]{schreiber2009aqft}
Urs Schreiber.
\newblock {AQFT} from $n$-functorial {QFT}.
\newblock {\em Communications in Mathematical Physics}, 291(2):357--401, 2009.

\bibitem[Seg88]{segal1988definition}
Graeme~B Segal.
\newblock The definition of conformal field theory.
\newblock In {\em Differential Geometrical Methods in Theoretical Physics},
  pages 165--171. Springer, 1988.

\end{thebibliography}

\end{document}